# Reducing the Risk of Spreadsheet Usage – a Case Study

Mel Glass (EASA), David Ford (Amlin), Sebastian Dewhurst (EASA)
Davenport House, 39 Evenlode Drive
Wallingford, Oxfordshire, OX10 7NZ, UK.
info@easasoftware.com

**ABSTRACT**

*The frequency with which spreadsheets are used and the associated risk is well known. Many tools and techniques have been developed which help reduce risks associate with creating and maintaining spreadsheet. However, little consideration has been given to reducing the risks of routine usage by the "consumers"- for example when entering and editing data. EASA's solution, available commercially, ensures that any routine process involving spreadsheets can be executed rapidly and without errors by the end-users, often with a significant reduction in manual effort. Specifically, the technology enables the rapid creation and deployment of web-based applications, connected to one or more centralized spreadsheets; this ensures version control, easy and error-free usage, and security of intellectual property contained in spreadsheets.*

## 1. INTRODUCTION

Countless surveys have revealed the ubiquitous nature of spreadsheets; it has even been suggested that spreadsheets are a universal technology [Durfee, 2004].

This is especially true in the financial sector, due in part to the relative extremes of specialisation within the industry [Sentence, 2006]. Specialisation of end-user knowledge, coupled with short delivery timeframes, makes it difficult to deliver custom solutions using conventional IT technologies. It is simply quicker for end-users to build their own tools with spreadsheets than it is to explain to an IT professional what kind of application is required, and then wait for it to be built.

This, of course, has given rise to the recognition of the serious lack of spreadsheet control [Davies & Ikin, 1987; Cragg & King, 1993; Fernandez, 2002; Floyd, Walls, & Marr, 1995; Gosling, 2003; Hall, 1996; Hendry & Green, 1994; Nardi & Miller, 1991; Nardi, 1993; Schultheis & Sumner, 1994].

Methodologies and commercial products intended to identify errors in spreadsheets during the creation, testing, and subsequent maintenance are becoming more common-place [Ayalew et al, 2000; Bekenn, 2008]. However, little consideration has been given to control of the spreadsheets as they are used by the consumers who use these spreadsheets to execute routine processes. There are some notable exceptions [Chambers & Hamill, 2008], and in fact some have considered the Excel "supply chain" [Samar & Patni, 2005]. The all-to-frequent result of this is that, if the problem is addressed at all, extreme measures are implemented in an effort to control usage; for example, we see companies who prevent changes to "approved" spreadsheets by simply locking the spreadsheet beyond input values, which is moderately effective but which also has limitations [Baxter, 2007].

We also see the wholesale replacement of operational spreadsheets with custom applications. While this is a preferred route, it is usually expensive and time consuming; therefore spreadsheets will always fill the void between what a business needs today and the formal installed systems, as noted by PWC [Baxter, 2007].

## 2. AN ALTERNATIVE APPROACH

Our approach, available as a commercially software package called EASA (Enterprise Accessible Software Applications), is a tool for building Web-based applications. These can run existing spreadsheets (and other software) on a Web server, thereby improving usability, accessibility, and control.

### 2.1 Overview of EASA

EASA has been in use since 2002, by industries as diverse as health-care, communications, energy, financial services, and manufacturing. Specific uses have included the provision of custom interfaces to multiple existing applications, giving users simplified access to key software tools and data [Kornfein & Rajiv, 2008]. Another common purpose is the modernization of legacy software, which can be “wrapped” by EASA and transformed into modern, web-enabled applications, accessible to any desktop, laptop, or mobile device in the enterprise [Casanova, 2003]. EASA’s architecture is shown in Figure 1.

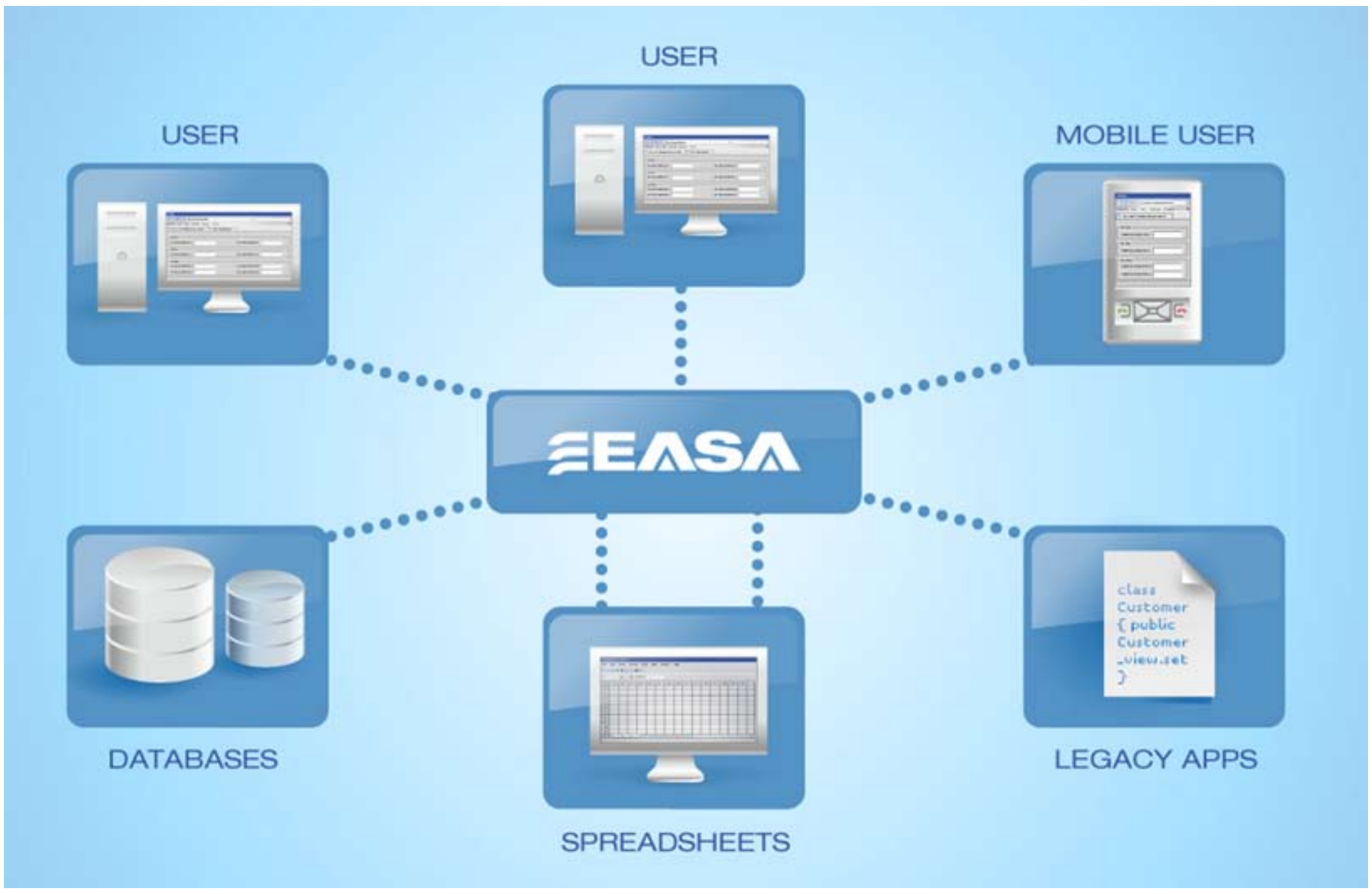

**Figure 1. EASA architecture**

However, the area of present interest is the application of the EASA product to spreadsheets; this approach allows the creation of simple web based applications (or EASAPs) that link to one or more key spreadsheets, databases, and other existing software. This not only eliminates the need to distribute key spreadsheets, but also ensures that they are used *precisely* the way their authors intended, preventing almost all user-errors. A typical EASAP is shown in Figure 2.

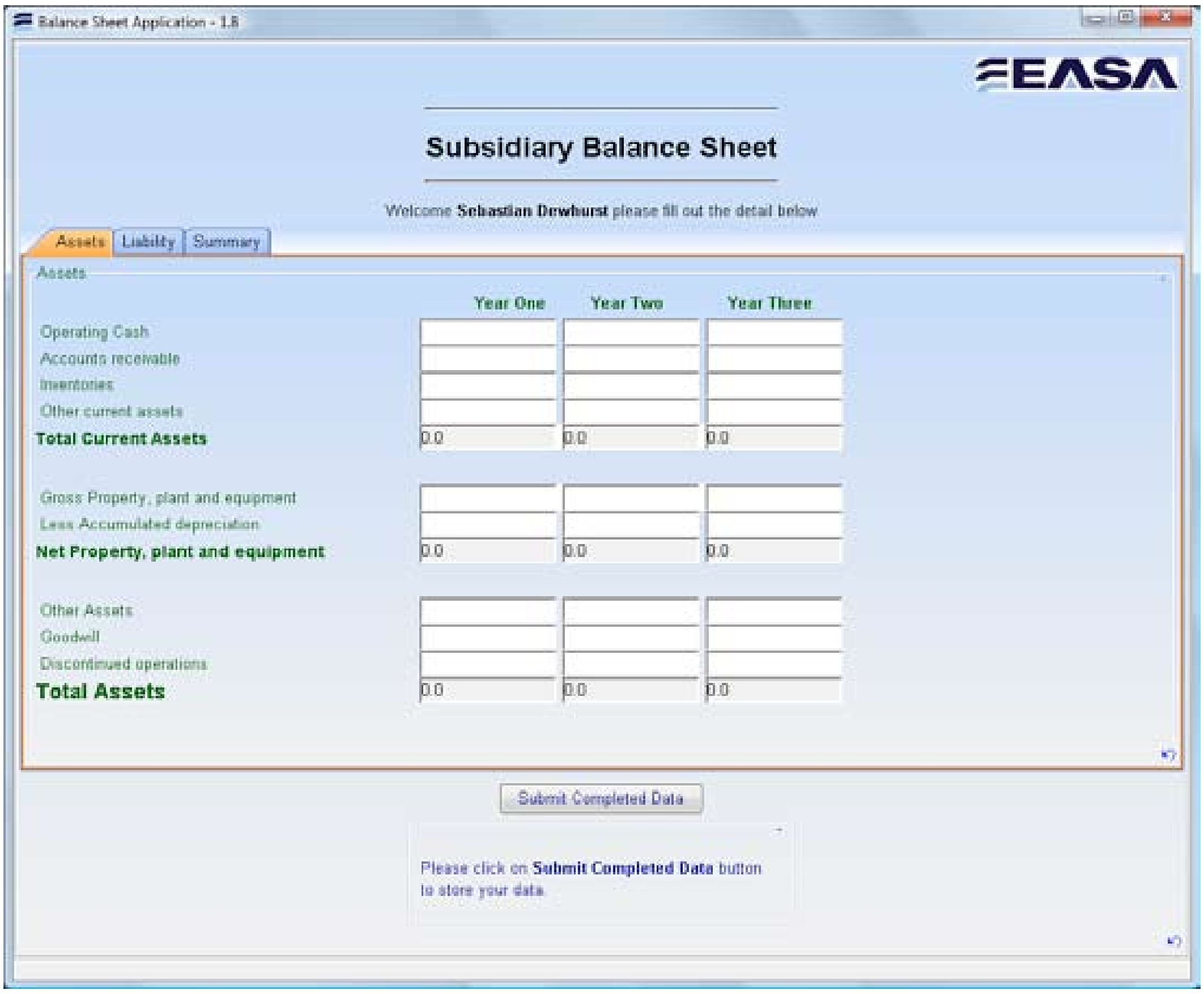

**Figure 2. A typical EASA application, or "EASAP", which connects to a central spreadsheet**

A side benefit includes the ability to link spreadsheets with other enterprise software, and to provide user-specific views – only relevant information is exposed to users. Finally, it is possible to establish a record of who did what, with which spreadsheet, and when.

The results pages provide an option to share reports on completed work with other users; their content is defined by the author, and can contain text, tables, charts, images or even animations generated by the application. Figure 3 shows a typical example.

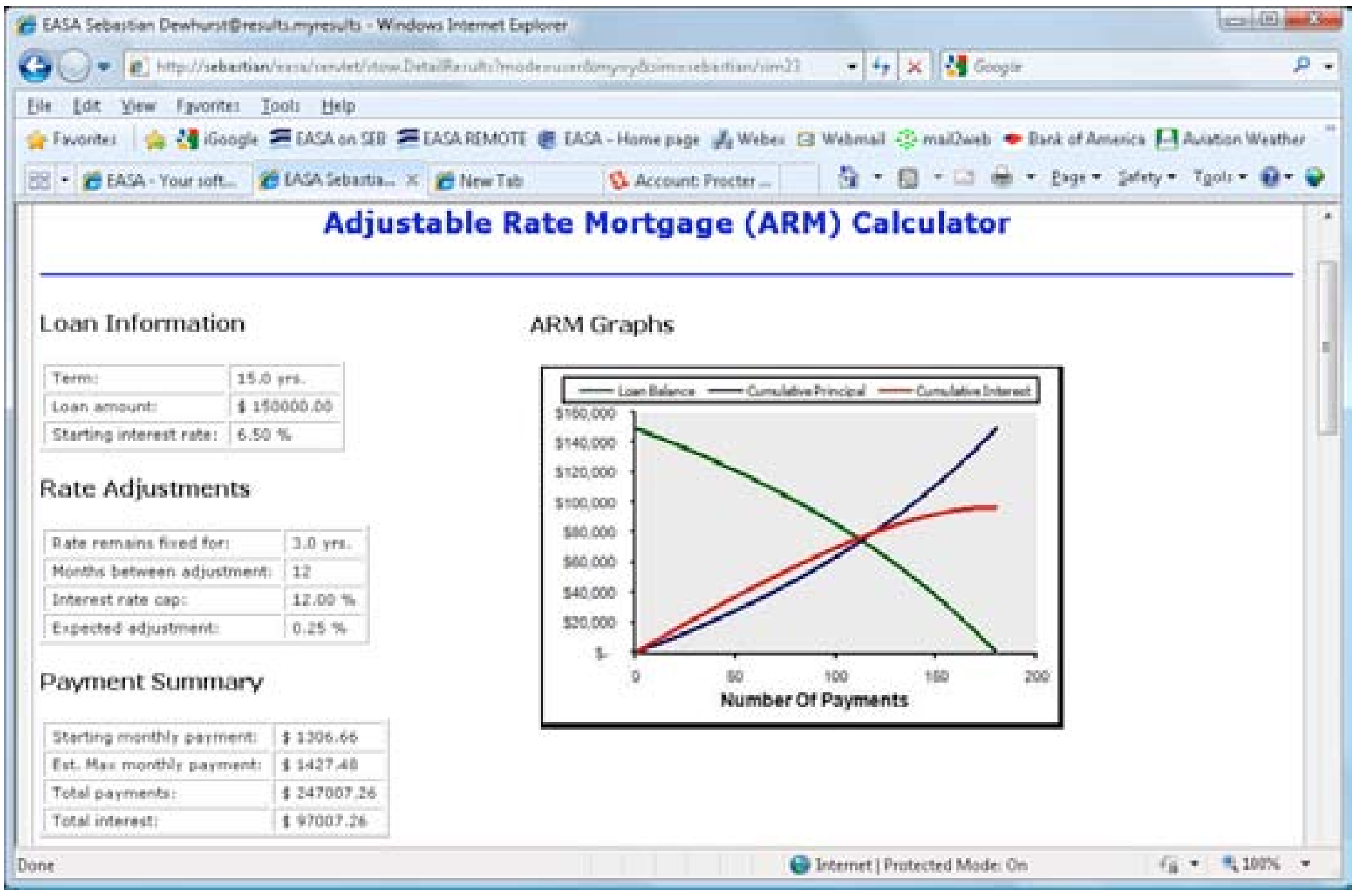

**Figure 3. Example Report Page from EASA**

### 2.2 What is EASA?

EASA permits the rapid creation and deployment of custom, web-enabled tools without the need for skilled programmers, helping organizations save time and money by improving the efficiency of their processes. EASA applications, or EASAPs, can be deployed over a corporate intranet or over the internet.

A custom EASAP (EASA Application) can drive multiple underlying software tools installed on existing systems throughout the enterprise. The underlying software may be anything from complex "expert only" applications, to legacy systems, modern databases, and spreadsheets.

Each EASAP is built for a specific need within the organization. EASAPs are available over the intranet to authorized users throughout the enterprise, providing secure, simplified access to the company's processes, best practices, expertise and software assets.

EASA's codeless application builder is used to create a user interface, to link the interface to underlying software and databases, and if required, to create custom reports. This eliminates complicated and time-consuming coding of custom applications with tools such as C++, VB, and Java and their associated Integrated Development Environments, which might take weeks, months, or even longer. By comparison, EASA allows new custom applications to be created, tested, and deployed in as little as a few hours.

### 2.3 How does EASA work?

The EASA software is a client-server environment, in which the information needed to run a particular software application is locked into an EASAP by the author. An EASAP can drive any batch-capable or COM-compliant software (and has dedicated wizards for linking to Excel), and also allows interaction with ODBC compliant databases. EASA contains the necessary structures to run applications on different computers (where they are resident), to control queuing and user access, and to serve up web-pages to the users.

### 2.4 How does the Author create an EASAP?

The authoring tools are a particularly powerful part of the EASA system, and a typical EASAP can be created in 2 to 8 hours. EASAP Builder provides a tree structure for building the EASAP and linking it to a centrally maintained spreadsheet, databases and/or other software assets. No knowledge of programming is needed. Figure 4 shows an example of an author's tree for a typical EASAP.

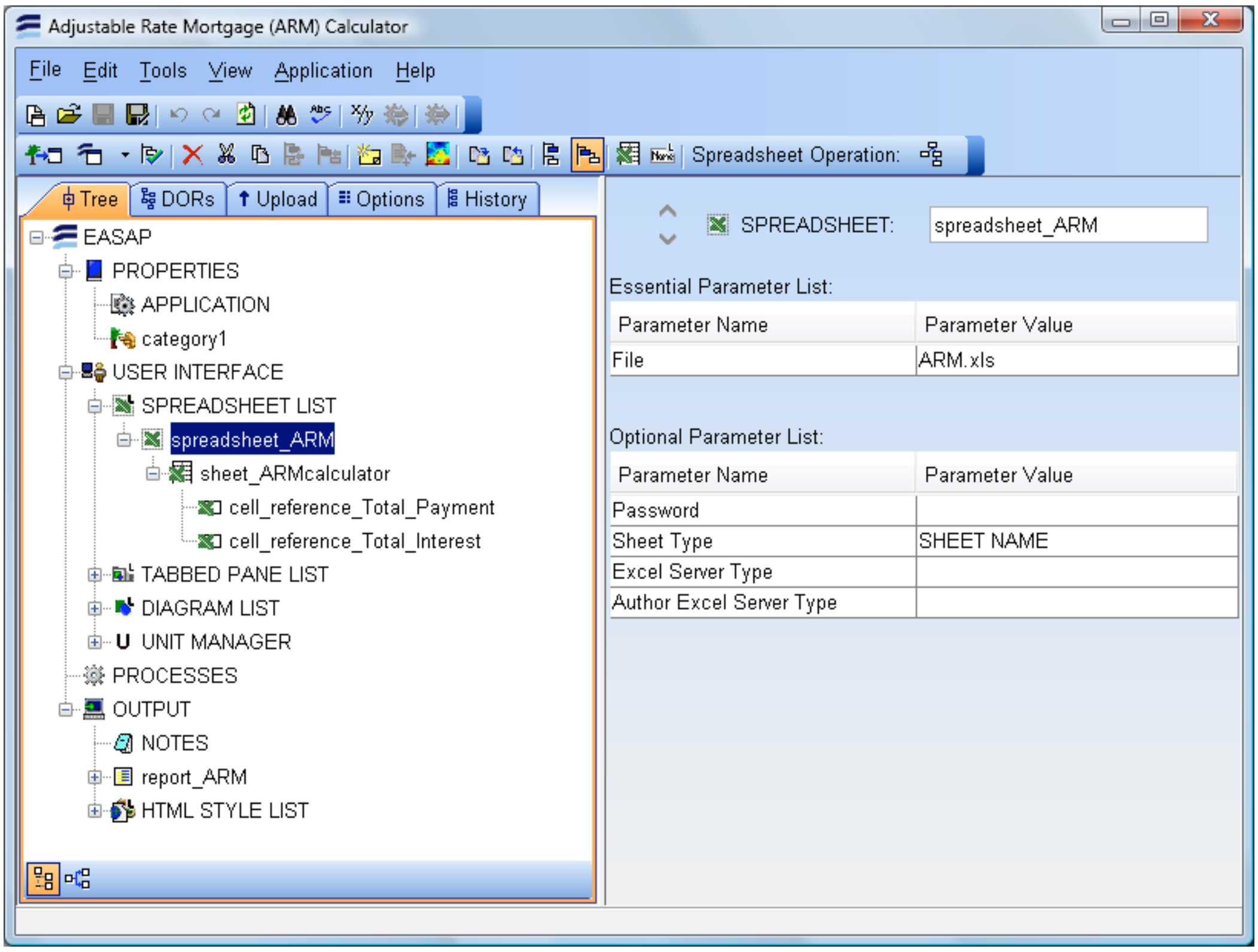

**Figure 4. EASAP Builder – typical screen**

EASAP Builder allows an author to create:
(a) user-interface components such as tabbed panes, choice lists, radio buttons, and input fields;
(b) components that validate data typed into the input fields;
(c) components that perform calculations from this by passing it to a spreadsheet (or other programs);
(d) user-interface components that display results.

EASA also provides a version control system for the EASAPS – allowing them to be developed "off line" by an Author, then, when ready, to publish them on the Intranet. Any subsequent versions are then automatically given revision numbers. Previous versions of EASAPS can be restored by the system administrator and EASAPs can be automatically upgraded to run with the latest version of EASA.

## 2.5 Key differences between EASA and Sharepoint's Excel Services

EASA is not the only software for running spreadsheets over the Web: one can also do so with Microsoft's Excel Server. However, Excel Server has two drawbacks:

Firstly, the user experiences an interface that must look like a spreadsheet, while using EASA, an author can design a more intuitive interface, so that users are less likely to make mistakes.

Secondly, if a critical spreadsheet has add-ins or macros, it will not work under Excel Services, while it will under EASA. This is no small issue – we have seen companies list the spreadsheets they wish to deploy, only to find that many hundreds of them have add-ins which would require rework of the order of weeks or even months per spreadsheet. We summarise key differences in Figure 5 below.

| Feature | MS Performance Point | MS Excel Services SharePoint | EASA |
|---|---|---|---|
| Tailored productivity tools which create web applications specific to individual enterprises' requirements. | No | No | Yes |
| Database neutral with any SQL database being equally easy to set up. Including all the internal data which can build up dependency to unwanted database technologies. | No | No | Yes |
| Simple installation and configuration. | No | No | Yes |
| Automatic database backing and storage of data to predesigned tables. | Yes | No | Yes |
| Codeless custom web application building linked to Excel Service automatically. | No | No | Yes |
| Data aggregation in database and automatically linked to template sheet for report generation. | No | No | Yes |
| Possible to aggregate and link several spreadsheets in one web application | ? | No | Yes |
| Extensible using Java® Technology with extensions being reusable between applications and servers. | No | No | Yes |
| "One version of the truth" design for deployment of spreadsheets to the enterprise. | No | Yes | Yes |
| Desktop Excel not Required | No | No | Yes |
| Excel .xls file can be downloaded with current working data in and used locally for reporting. | Yes | Yes | Yes |
| Product neutral. | No | No | Yes |
| Avoidance of Excel client configuration for whole enterprise (i.e. reduced IT overhead) | No | No | Yes |
| Simple and robust security system | No | No | Yes |
| Reusable custom components which can be used and reused by non-programmers. | No | No | Yes |

**Figure 5. Key differences between EASA and Sharepoint's Excel Services**

## 3. CASE STUDY

Amlin [www.amlin.com] is a leading independent insurer operating in the Lloyd's, UK, Europe and Bermudian markets. Amlin provides insurance cover to commercial enterprises and reinsurance protection to other insurance companies around the world.

As in any organization, many key processes at Amlin are under-pinned by spreadsheets; spreadsheets are easy to create, familiar to users, and flexible. However, lack of control means that the risk of mistakes and the cost of executing the process are unacceptably high.

### 3.1 Amlin’s RDS application

Staff at Amlin identified one process particularly in need of better control – RDS, or Realistic Disaster Scenario.

RDS requires every agent to submit a six monthly report to Lloyd’s, showing exposure against scenario and by insurance risk code. In Amlin’s case, this means some 20 underwriters around the world must each submit reports.

Historically, this process has been spreadsheet-based; AMLIN underwriters complete and submit a spreadsheet cloned from a template. Complicating this process is the fact that the data required by Lloyd’s changes on a regular basis. Hence a new template must be created, tested and distributed each time there is a change.

The issues which arise include:

- Users change the spreadsheet to suit local requirements before completing and submitting, which results in aggregation errors down-stream.
- The spreadsheet must serve every user, and is therefore more complex than any one user needs – and most users have access to much more of the spreadsheet than is necessary.
- Any change requires that the new spreadsheet is distributed to every user – with instructions to remove the out-dated version. There is no mechanism to ensure the correct version is being used.
- Aggregation is carried out using a series of linked spreadsheets. Links have to be constantly updated, and many layers of spreadsheet refreshed.
- The data exists in many spreadsheets and not in a database, so historical comparison is time-consuming and expensive.

The original spreadsheet is shown in Figure 6.

| SCENARIO | GULF OF MEXICO HURRICANE (Texas only - Onshore Property Loss $95bn - Offshore Energy Loss $11bn) - (GOM) | | | | | | | | | | AS AT : | 10/1/2007 |
|---|---|---|---|---|---|---|---|---|---|---|---|---|
| | AIR - EVENT ID 270005783, INDUSTRY INSURABLE LOSS 93,324,226,000. | | | | | | | | | | | |
| DESCRIPTION | RMS - EVENT ID 44103, INDUSTRY INSURABLE LOSS 97,905,471,489 USD. | | | | | | | | | | | |
| ACCOUNT CLASS | MARINE ENERGY | | | | | | | | | | CCY | USD |
| TYPE OF BUSINESS | LLOYD'S CATEGORY | MODELLING TYPE | TRUST FUND | SETT CCY | LONDON TOTAL GROSS EXPOSED AGGREGATE | BERMUDA CESSION AGGREGATE | LONDON NET EXPOSED AGGREGATE | Damage Ratio % | LONDON GROSS LOSS | LONDON NET LOSS | FAC R/I RECOVERIES | |
| COW | Energy Property Damage | Modelled Internally | SL | USD | $590,992,119 | $143,348,659 | $447,643,460 | 7.85% | $45,775,118 | $35,156,503 | | |
| ICR | Energy Property Damage | Modelled Internally | SL | USD | $4,500,000 | $1,125,000 | $3,375,000 | 22.65% | $1,019,250 | $764,438 | | |
| LOPI | Energy Property Damage | Modelled Internally | SL | USD | $2,129,059 | $532,265 | $1,596,794 | 97.13% | $2,068,030 | $1,551,022 | | |
| PD | Energy Property Damage | Modelled Internally | SL | USD | $140,211,831 | $26,520,419 | $113,691,412 | 14.70% | $19,508,468 | $16,713,434 | | |
| ROD | Energy Property Damage | Modelled Internally | SL | USD | $36,461,961 | $7,565,493 | $28,896,468 | 2.17% | $792,399 | $625,784 | | |
| | | | | | | | $0 | #DIV/0! | | | | |
| | | | | | | | $0 | #DIV/0! | | | | |
| | | | | | | | $0 | #DIV/0! | | | | |
| | | | | | | | $0 | #DIV/0! | | | | |
| | | | | | | | $0 | #DIV/0! | | | | |
| | | | | | | | $0 | #DIV/0! | | | | |
| | | | | | | | $0 | #DIV/0! | | | | |
| | | | | | | | $0 | #DIV/0! | | | | |
| | | | | | | | $0 | #DIV/0! | | | | |
| | | | | | | | $0 | #DIV/0! | | | | |
| | | | | | | | $0 | #DIV/0! | | | | |
| | | | | | | | $0 | #DIV/0! | | | | |
| | | | | | | | $0 | #DIV/0! | | | | |
| | | | | | | | $0 | #DIV/0! | | | | |
| TOTAL | | | | | $774,294,970 | $179,091,836 | $595,203,134 | 9.21% | $69,163,265 | $54,811,181 | $0 | |

**Figure 6. Before: Amlin's Realistic Disaster Scenario reporting process required many users to complete and submit a complex Excel spreadsheet. Manual aggregation was time-consuming and error-prone**

## 3.2 Reducing the risk and cost

A solution requiring the complete elimination of Excel was considered. However, that would have required significant investment in building a database application, discarding all the business intelligence already contained in the spreadsheet.

Instead, EASA's spreadsheet management solution offered a far more cost-effective alternative, allowing Amlin to secure a master version of the spreadsheet on a server. Users now access it only via a custom web application created with EASA's codeless application builder, allowing a more natural work-flow.

- The custom web application is so intuitive that training is no longer required (see Figure 7).
- Users only see what they need to see; they no longer access the spreadsheet directly, and are not able to "dabble" with it.
- If a change is required, it is made in one place and is immediately published to all users; version control is ensured.
- Aggregation is now automatic

The RDS application is now used in the UK, Bermuda and Singapore. The RDS return for Lloyd's is now produced at the push of a button, and the process has been error-free since the deployment with EASA.

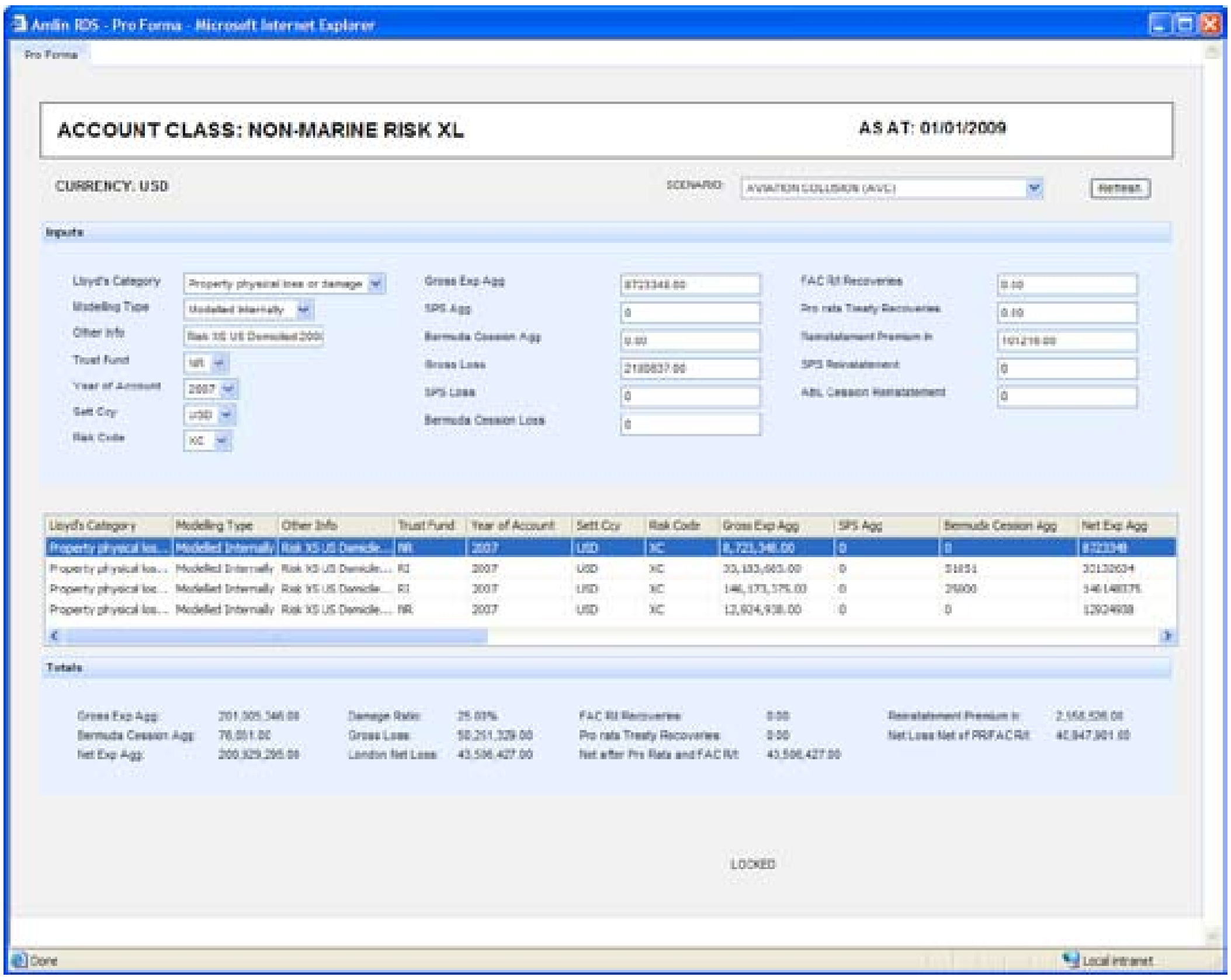

**Figure 7. After: the new web-based RDS application, created with EASA's codeless application builder, is far easier to deploy and use. It leverages the existing spreadsheet and the business intelligence already embedded in it, eliminates usage errors, and ensures version control**

## 4. CONCLUSION

A new approach to control the use of critical spreadsheets has been presented, along with a case study demonstrating the effectiveness of the approach. The existing spreadsheets in a company can be leveraged without re-work.

Because this approach facilitates only the deployment and usage of critical spreadsheets, it is complementary to other methods and products which ensure the integrity and auditability of the master spreadsheet itself.

Finally, the technology used allows for integration of critical spreadsheets with other software systems (even legacy applications), and can also be used to provide access to critical spreadsheets for mobile users.